\documentclass[aps,prc,amssymb,amsmath,twocolumn,superscriptaddress]{revtex4}

\usepackage[english]{babel}
\usepackage[utf8]{inputenc}

\usepackage[dvips]{graphicx}
\usepackage[sort&compress]{natbib}

\begin{document}

\title{Conical Emission from Shock Waves in Ne(1--20~AGeV)+U
  Collisions}

\author{Philip Rau}%
\email{rau@th.physik.uni-frankfurt.de}%
\affiliation{Institut f\"ur Theoretische Physik, Johann Wolfgang
  Goethe-Universit\"at, Max-von-Laue-Str.~1, 60438 Frankfurt am Main,
  Germany}%
\author{Jan Steinheimer}%
\affiliation{Institut f\"ur Theoretische Physik, Johann Wolfgang
  Goethe-Universit\"at, Max-von-Laue-Str.~1, 60438 Frankfurt am Main,
  Germany}%
\author{Barbara Betz}%
\affiliation{Institut f\"ur Theoretische Physik, Johann Wolfgang
  Goethe-Universit\"at, Max-von-Laue-Str.~1, 60438 Frankfurt am Main,
  Germany}%
\affiliation{Department of Physics, Columbia University, 538 West
  120th Street, New York, 10027, USA}%
\author{Hannah Petersen}%
\affiliation{Institut f\"ur Theoretische Physik, Johann Wolfgang
  Goethe-Universit\"at, Max-von-Laue-Str.~1, 60438 Frankfurt am Main,
  Germany}%
\affiliation{Frankfurt Institute for Advanced Studies (FIAS),
  Ruth-Moufang-Str.~1, 60438 Frankfurt am Main, Germany}%
\affiliation{Department of Physics, Duke University, Durham, North
  Carolina 27708-0305, USA}%
\author{Marcus Bleicher}%
\affiliation{Frankfurt Institute for Advanced Studies (FIAS),
  Ruth-Moufang-Str.~1, 60438 Frankfurt am Main, Germany}%
\author{Horst St\"ocker}%
\affiliation{Institut f\"ur Theoretische Physik, Johann Wolfgang
  Goethe-Universit\"at, Max-von-Laue-Str.~1, 60438 Frankfurt am Main,
  Germany}%
\affiliation{GSI Helmholtzzentrum f\"ur Schwerionenforschung GmbH,
  Planckstr.~1, 64291 Darmstadt, Germany}%

\begin{abstract}
  The formation and propagation of high-density compression waves,
  e.g.\ Mach shock waves, in cold nuclear matter is studied by
  simulating high-energy nucleus-nucleus collisions of Ne with U in
  the energy range from $E_{\rm lab} = 0.5$~AGeV to $20$~AGeV. In an
  ideal hydrodynamic approach, the high-density shock wave created by
  the small Ne nucleus passing through the heavy U nucleus is followed
  by a slower and more dilute Mach shock wave which causes conical
  emission of particles at the Mach cone angle. The conical emission
  originates from low-density regions with a small flow velocity
  comparable to the speed of sound.  Moreover, it is shown that the
  angular distributions of emitted baryons clearly distinguish between
  a hydrodynamic approach and binary cascade processes used in the
  Ultra-relativistic Quantum Molecular Dynamics (UrQMD) transport
  model.
\end{abstract}

\maketitle
\section{Introduction}
\label{sec:introduction}
Hydrodynamic models predict a sideward emission of nuclear matter in
fast nucleus-nucleus collisions due to the transformation of kinetic
energy of the projectile into compression and heat energy of the
medium.  High-density shock waves created during the collision of
unequal nuclei push matter in transverse direction, generating a
measurable preferential emission at a well-defined Mach
angle~\cite{Hofmann:1975by,Baumgardt:1975qv,Hofmann:1976dy,Stoecker:1977cn,Stoecker:1980vf,Stocker:1981zz}.
This angle is connected to the medium properties, in particular to the
velocity of sound
\begin{equation}
  \label{eq:sound-velocity}
  c_{\rm s}^2 = \partial p/\partial e\text{,}
\end{equation}
by the classical Mach cone formula
\begin{equation}
  \label{eq:mach-angle}
  \theta_{\rm MC} =  \cos^{-1}(c_{\rm s}/v_{\rm sh})\text{.}
\end{equation}
Here, $v_{\rm sh}$ denotes the velocity of the leading head shock wave
traveling through the target nucleus.\par
In recent years, the investigation of (Mach) shock waves has regained
tremendous attention, both on the
theoretical~\cite{Stoecker:2004qu,Satarov:2005mv,Ruppert:2005uz,CasalderreySolana:2004qm,CasalderreySolana:2006sq,Chaudhuri:2005vc,Renk:2005si,Stoecker:2007su,Neufeld:2008fi,Betz:2008wy,Betz:2008ka,Noronha:2008,Bouras:2009nn}
as well as on the experimental
side~\cite{Adams:2005ph,Adler:2005ee,Adler:2005ad,Ulery:2005cc,Ajitanand:2006is,Ulery:2007zb,Ulery:2008pj,:2008cqb,abelev:052302},
considering shock waves from partonic projectiles.\par
In this paper, we return to the original idea and re-investigate the
creation of Mach-like shock waves as well as the sideward deflection
of matter in asymmetric nucleus-nucleus collisions using
(3+1)-dimensional ideal hydrodynamics~\cite{Rischke:1995pe} and the
Ultra-relativistic Quantum Molecular Dynamics (UrQMD) transport
model~\cite{Bass:1998ca,Bleicher:1999xi}. The conical emission at Mach
angles $\theta_{\rm MC}$ (where $\theta$ denotes the polar angle
between the beam axis and the flux of matter) persists over a wide
range of beam energies and impact parameters, allowing for a
comprehensive study even at lower beam energies. We analyze the change
of the particle emission angle $\theta_{\rm lab}$ with increasing
impact parameter and show the dependence of the Mach shock wave on the
size of the projectile nucleus.\par
The (3+1)-dimensional ideal hydrodynamic calculations are performed
for central and non-central Ne+U collisions at $0.5$--$20$~AGeV,
applying a chiral hadronic equation of state (EoS) that exhibits a
phase transition to a chirally restored phase, as it is expected by
lattice calculations
\cite{Fodor:2001pe,Karsch:2004wd,Fodor:2007vv}. Therefore, we are able
to investigate the impact of this phase transition on the emission
angle of particles from Mach-like shock waves in heavy ion
collisions.\par
Additionally, analogous calculations are performed using the UrQMD
transport model. The comparison of the different models with
experimental data will allow to draw conclusions about the underlying
process of the conical emission. While the conical emission of the
reaction products in hydrodynamics is due to the formation of shock
waves in the nuclear matter, the emission pattern in the UrQMD model
is generated by binary collisions and does not show a Mach cone
pattern.

\section{Hydrodynamics}
\label{sec:hydrodynamics}
Each hydrodynamic simulation of a collision process is initialized
when the Ne nucleus starts to penetrate the U nucleus and is performed
in the rest frame of the Uranium target. The ground state energy
density distribution of each nucleus is given by the Woods-Saxon form
\begin{equation}
  \label{eq:woods-saxon}
  e(r) = e_0/\left[ 1 + e^{\left( r-r_0 \right)/d}\right] \text{,}
\end{equation}
where $e_0 = 147$~MeV/fm$^3$ denotes the ground-state energy
density.\par
The mean radius of a nucleus of mass $A$ is computed via $r_0 = \left(
  1.12\,A^{(1/3)} - 0.86\,A^{(-1/3)} \right)$~fm and the surface
thickness is set to $d = 0.54$~fm~\cite{DeVries1987495}. In the
longitudinal direction the density distribution of the projectile is
Lorentz contracted.  The distribution of the net baryon number density
in the cold nuclei is computed accordingly with $n_0 =
0.16$~fm$^{-3}$.\par
Ideal hydrodynamics denotes the conservation of energy, momentum, and
net baryon number
\begin{align}
  \label{eq:conservation-eqns}
  \begin{split}
    \partial_{\mu} T^{\mu \nu} &= 0 \text{,}\\
    \partial_{\mu} \left( n u^{\mu} \right)&= 0.
  \end{split}
\end{align}
These equations are numerically solved by the SHASTA
algorithm~\cite{Boris197338,Book1975248,Rischke:1995ir,Rischke:1995mt,Rischke:1995pe}
on a discretized grid with the ideal hydrodynamic energy-momentum
tensor given by $T^{\mu \nu} = (e + p) u^{\mu}u^{\nu} - pg^{\mu\nu}$
(with the four-velocity of the fluid $u^{\mu} =
\gamma\,(1,\mathbf{v})$, $\gamma = (1-v^2)^{-1/2}$, and the net baryon
density $n$). The algorithm ensures the flux-corrected transport of
the thermodynamic quantities over the computational grid as described
in Refs.~\cite{Rischke:1995ir,Boris197338,Book1975248}.  The symmetric
Eulerian grid is fixed in the target (computational) frame with a
static cell size $\Delta x = 0.2$~fm.  Time steps are set to $\Delta t
= 0.08$~fm. Changing these numbers does not significantly affect the
results.\par
\begin{figure}[t]
  \centering 
  \includegraphics[width=1.\columnwidth]{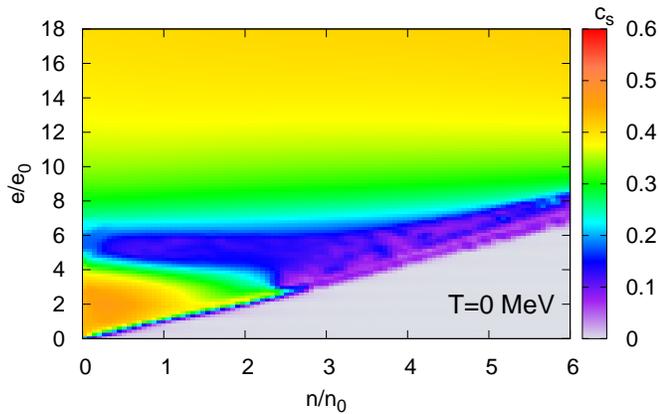}
  \caption{(Color online) The sound velocity $c_s$ of the EoS used in
    our calculations as a function of $e/e_0$ and $n/n_0$.}
  \label{fig:EoS-map}
\end{figure}
The pressure $p$ is connected to the energy density $e$ and the net
baryon number density $n$ by the EoS $p(e,n)$.  For this study a
hadronic EoS, derived from a chiral hadronic SU(3) Lagrangian which
includes the lowest baryon octet together with the multiplets of
scalar, pseudo-scalar, vector, and axial-vector mesons, is
used~\cite{Steinheimer:2007iy}. All parameters of the model EoS are
fixed either by symmetry relations, hadronic vacuum observables or
nuclear matter saturation properties. The model exhibits a smooth
decrease of the chiral condensates (cross over) for high temperature
and low baryonic potential~\cite{Papazoglou:1998vr,Zschiesche:2001dx}.
In addition, the model also provides a satisfactory description of
realistic (finite-size and isospin asymmetric) nuclei and of neutron
stars~\cite{Papazoglou:1998vr,Schramm:2002xi,Schramm:2002xa}.
Additional baryonic degrees of freedom change the smooth cross over
into a first-order phase transition with a critical end point (CEP) at
$T \simeq 180$~MeV, $\mu_q \simeq 115$~MeV, depending on the
couplings~\cite{Theis:1984qc,Zschiesche:2001dx,Zschiesche:2004si,Zschiesche:2006rf,Steinheimer:2007iy}.
These values for the CEP are within the range expected from lattice
QCD
calculations~\cite{Fodor:2001pe,Fodor:2004nz,Fodor:2007vv,Karsch:2004wd}.
For a detailed discussion of this EoS see
Refs.~\cite{Zschiesche:2006rf,Steinheimer:2007iy}. The velocity of
sound $c_s$ in the $n$-$e$-plane of this EoS is depicted in
Fig.~\ref{fig:EoS-map}.\par
\begin{figure*}[t]
  \centering 
  \includegraphics[width=1.\textwidth]{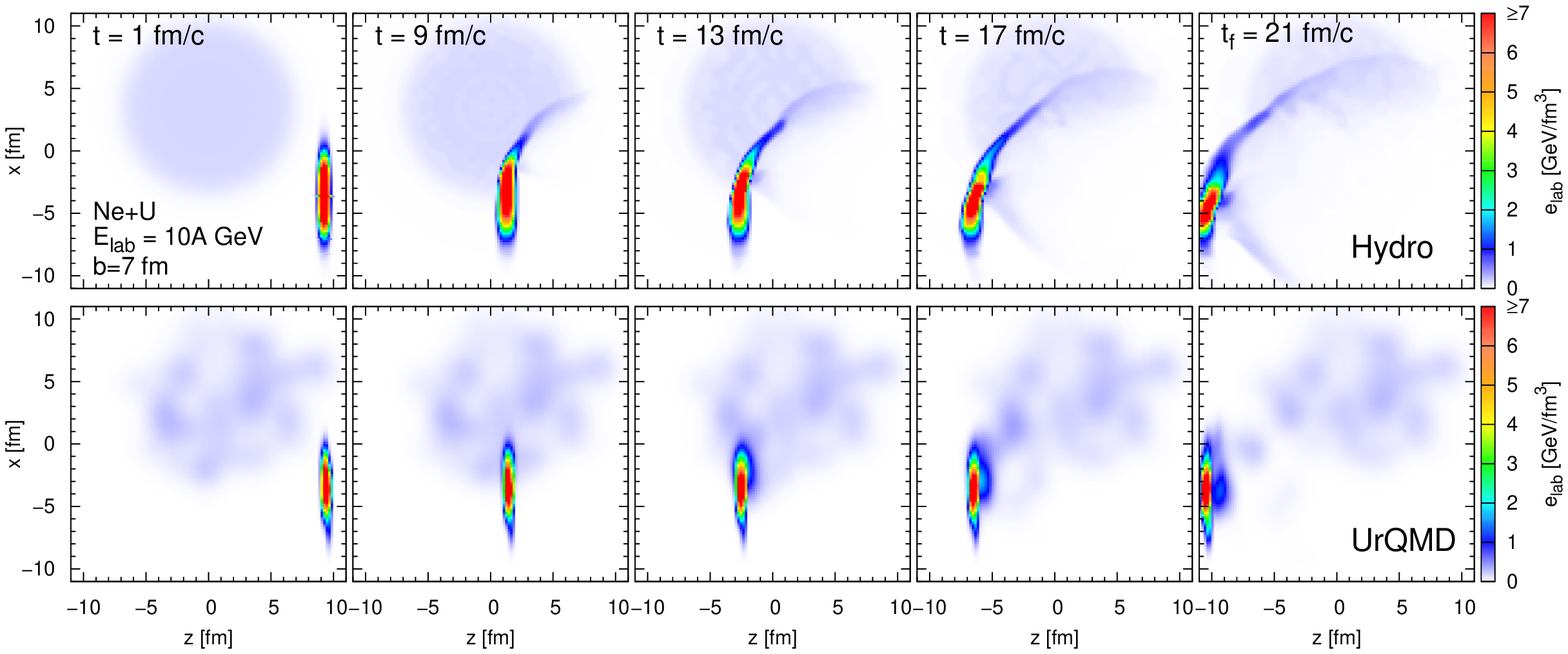}
  \caption{(Color online) Laboratory energy density ($e_{\rm lab}$)
    distribution in the reaction plane at different stages of a Ne+U
    collision at $E_{\rm lab} = 10$~AGeV and $b=7$~fm, calculated
    using the hydrodynamic (upper panel) and the UrQMD framework
    (lower panel) from initial to final state. Note that in this plot
    the laboratory energy density is cut off at $7$~GeV/fm$^3$.}
  \label{fig:time_evolution_cf_energy}
\end{figure*}
As the projectile nucleus hits the target with supersonic velocity,
the nuclear matter in the collision zone gets highly compressed and
thereby creates a strong shock wave that moves with $v_{\rm sh} \ge
c_s$ through the target nucleus. The shock velocity
\begin{equation}
  \label{eq:shock-velocity}
  v_{\rm sh} = \left[ \frac{(p_2 - p_1) (e_1 + p_2)}{(e_2 - e_1) (e_1 +
      p_1)} \right]^{1/2}
\end{equation}
can be derived from the one-dimensional relativistic shock adiabat
(\emph{Taub adiabat}~\cite{Taub:1948zz}).  Quantities with index 1
denote the unperturbed nuclear matter while the index 2 denotes the
energy and pressure in the compression zone. Unlike the
Glassgold-Heckroth-Watson approach~\cite{Glassgold:1959tu} and the
participant-spectator model~\cite{Westfall:1976fu}, in head-on
collisions the evolving violent shock wave completely destroys the
entire target nucleus, leaving no target fragments in the target frame
at all. The very hot, dense, and fast moving region near the collision
axis is commonly referred to as \emph{head
  shock}~\cite{Stoecker:1980vf,Hofmann:1976dy,Hofmann:1975by,Betz:2008wy}. The
center of the head shock reaches the following maximal local rest
frame (lrf) energy and baryon number densities during the early
evolution of the shock wave
\begin{center}
  \begin{tabular}[t]{lcccc}
    beam energy: $E_{\rm lab}$~[AGeV] & 0.5 & 1 & 10 & 20\\
    \hline
    $e_{\rm max}/e_0$ (lrf) & 3.5 & 13 & 32 & 50\\
    $n_{\rm max}/n_0$ (lrf) & 3   & 10 & 11 & 11.
  \end{tabular}  
\end{center}
The further evolution is accompanied by a flow of energy to the outer
regions of the system. This causes a significant weakening of the head
shock.  For beam energies $E_{\rm lab} \ge 1$~AGeV the chirally
restored phase is reached in the head shock. At the back of the shock
wave, a slowly propagating region ($v \sim c_s$) with low energy
density is formed. In the head shock, matter is pushed ahead while it
gets deflected sidewards in the outer regions of the shock wave,
leading to a Mach shock structure in the
medium~\cite{Baumgardt:1975qv,Hofmann:1976dy}.\par
Figure~\ref{fig:time_evolution_cf_energy} shows the distribution of
the laboratory frame energy density $e_{lab}$ in the reaction plane at
different stages of a non-central ($b=7$~fm) collision of Ne+U at
$E_{\rm lab} = 10$~AGeV. The upper panel displays the collision
process in hydrodynamics, the lower panel in the UrQMD framework. In
hydrodynamics, the evolving shock wave is clearly visible. Note that
the laboratory energy density is cut off at $7$~GeV/fm$^3$ in this
figure.\par
The described shock wave creation is comparable to the scenario of a
fast parton jet which distributes energy and momentum to the
medium~\cite{Stoecker:2004qu,Satarov:2005mv,Ruppert:2005uz,CasalderreySolana:2004qm,CasalderreySolana:2006sq,Chaudhuri:2005vc,Renk:2005si,Stoecker:2007su,Neufeld:2008fi,Betz:2008wy,Betz:2008ka,Noronha:2008,Bouras:2009nn}.
Particles emitted directly from the head shock have large momenta in
forward direction and therefore are emitted at small $\theta_{\rm
  lab}$ angles, while particles emitted from outer regions of the
(Mach) shock wave have low momenta and are predominantly emitted in
sideward directions.\par
\begin{figure}[t]
  \centering
  \includegraphics[width=1.\columnwidth]{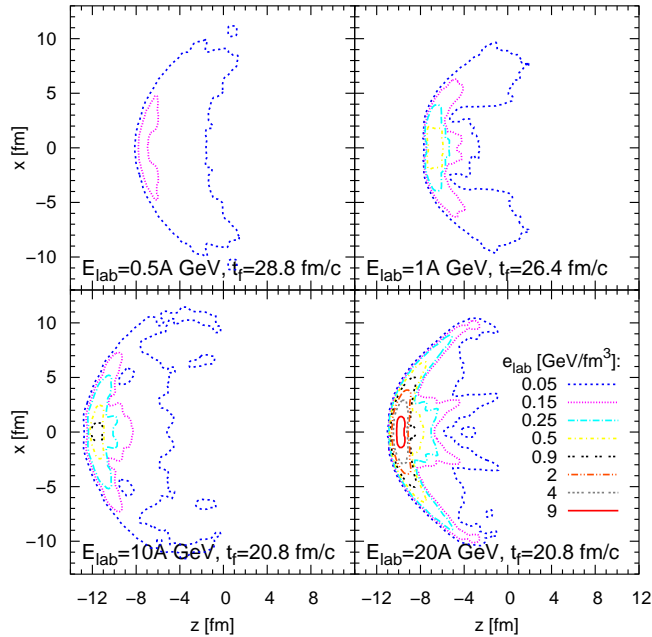}
  \caption{(Color online) Laboratory frame energy density $e_{\rm
      lab}$~[GeV/fm$^3$] distribution in the reaction plane at $t_f$,
    i.e.\ when the head shock wave has passed through the target
    nucleus, calculated in a hydrodynamic approach.  The maximum
    energy density in the center of the head shock is $e_{\rm lab}
    \approx 0.2,\, 1.6,\, 2,\, 11$~GeV/fm$^3$ for projectile energies
    of $E_{\rm lab} = 0.5,\, 1,\, 10,\, 20$~AGeV.}
  \label{fig:energy-maps}
\end{figure}

The hydrodynamic evolution is stopped at time $t_f$. Here, $t_f$ is
chosen large enough to ensure the full transition of the head shock
through the target nucleus. The energy density distribution in the
reaction plane in this final state $t_f$ is shown in
Fig.~\ref{fig:energy-maps} for central Ne+U collisions at different
beam energies, calculated using the hydrodynamic approach.  The shock
wave becomes stronger with increasing beam energy and thus increasing
energy density in the head shock. In fact, the energy density reaches
values up to $e_{\rm lab} = 11$~GeV/fm$^3$ at $t_f$ for the highest
beam energy of $E_{\rm lab} = 20$~AGeV (cf.\
Fig.~\ref{fig:energy-maps}).  Thus, the velocity of the head shock
comes close to $v_{\rm sh} = 1$ according to
Eq.~(\ref{eq:shock-velocity}).\par
The decoupling of the hydrodynamic system into particles is done at
$t_f$ performing an isochronous freeze-out. In a first scenario, we
consider a \emph{free streaming freeze-out} (FS FO). This is a rough
approximation for the distribution of light (He) and intermediate mass
(Li, Be) reaction products which exhibit much less ($\sim 1/A$)
thermal smearing than nucleons~\cite{Baumgardt:1975qv,Hofmann:1976dy}.
The spectra of the reaction products are obtained by computing the
relativistic kinetic energy of the nucleons from every cell on the
computational grid ($i$,$j$,$k$) via
\begin{equation}
  \label{eq:rel-kinetic-energy}
  E_{\rm kin}^{i,j,k} = \left( \gamma^{i,j,k} -1 \right) n^{i,j,k} M_N
\end{equation}
with a nucleon mass of $M_N = 939$~MeV and the Lorentz gamma factor
\begin{equation}
  \label{eq:gamma_factor}
  \gamma^{i,j,k} = \left[ 1 - \left( v_z^{i,j,k} \right)^2
  \right]^{-1/2} \text{,}
\end{equation}
where $v_z^{i,j,k}$ denotes the velocity and $n^{i,j,k}$ the net
baryon density of the specific fluid element. The polar angle between
the matter flux and the beam axis
\begin{equation}
  \label{eq:angle}
  \theta_{\rm lab} = \cos^{-1}\left( \frac{v_z^{i,j,k}}
    {\left|v^{i,j,k}\right|}
  \right)
\end{equation}
is computed for each cell and all baryons on the grid. I.e., thermal
smearing and resonance decays are neglected in this scenario.\par
A second freeze-out scenario is the Cooper-Frye
prescription~\cite{PhysRevD.10.186} (CF FO)
\begin{equation}
  \label{eq:cooper_frye}
  E\frac{dN}{d^3p} = \int d\sigma_{\mu}\; p^{\mu}f(x,p) 
  \text{,}
\end{equation}
where $f(x,p)$ denotes the boosted distribution function and
$d\sigma_{\mu} = (d^3x,\mathbf{0})$ the normal vector on the
hypersurface. We will consider the Cooper-Frye freeze-out with and
without final state interactions and resonance decays which are
performed using the UrQMD model as an afterburner. The details about
performing this Cooper-Frye freeze-out after the hydrodynamic
evolution as well as the integration into the UrQMD model are
described in detail in Ref.~\cite{Petersen:2008dd}. The hydrodynamic
calculations are contrasted with simulations within a binary collision
hadronic cascade, the default UrQMD model without a hydrodynamic
phase~\cite{Bass:1998ca,Bleicher:1999xi}. In the latter two approaches
the emission angle of the baryons with respect to the beam axis is
computed via $\theta_{\rm lab} = \cos^{-1} \left( p_z/|p|\right)$.\par
The hydrodynamic calculations show a sizeable fraction of particles
with low kinetic energies. They are emitted from regions with low flow
velocities behind the leading head shock at angles in the range of
$\theta_{\rm MC}$ (Eq.~(\ref{eq:mach-angle})). In contrast, particles
with high kinetic energies stemming from the highly compressed head
shock should be emitted under small angles. Since the head shock gets
remarkably stronger with increasing beam energy, an increasing number
of high energetic particles is expected at small angles and higher
beam energies.

\section{Results and Discussion}
\label{sec:results}
\begin{figure}[t]
  \centering
  \includegraphics[width=1.\columnwidth]{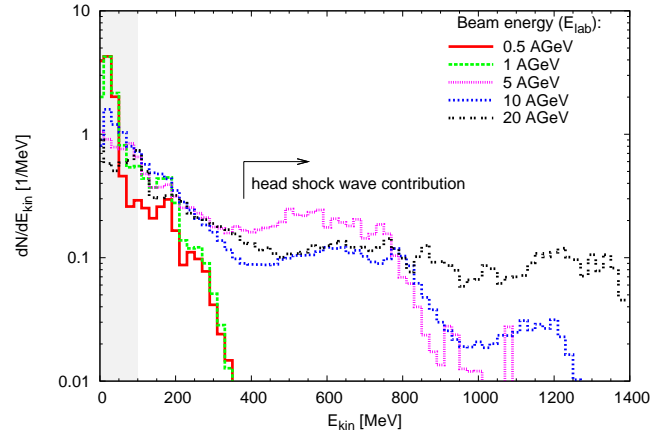}
  \caption{(Color online) Energy spectra of reaction products in Ne+U
    collisions. The majority of nucleons at the Mach angle arises from
    the low-energy (shaded) region $E_{\rm kin} \le 100$~MeV,
    corresponding to a nucleon velocity $v=0.43$. High-energy
    particles, however result from head shock regions.}
  \label{fig:results-energy-bins}
\end{figure}

Figure~\ref{fig:results-energy-bins} shows the kinetic energy
distributions of nucleons, obtained with the free streaming
freeze-out, for different beam energies.  Beam energies in the range
of $E_{\rm lab} \le 10$~AGeV show a distinct peak for nucleons with
low kinetic energies ($E_{\rm kin} \le 50$~MeV). For higher beam
energies, however, the peak height decreases, but due to the creation
of a stronger head shock more particles are produced at higher $E_{\rm
  kin}$.\par
\begin{figure}[t]
  \centering 
  \includegraphics[width=1.\columnwidth]{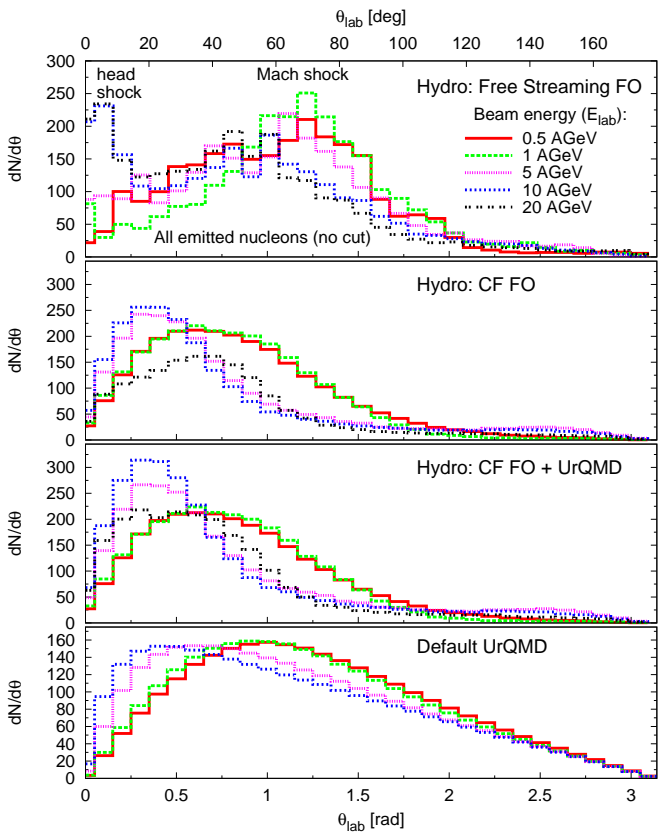}
  \caption{(Color online) Angular distributions of nucleons obtained
    from Ne+U reactions calculated using the hydrodynamic approach
    with the free streaming freeze-out (upper panel), the Cooper-Frye
    freeze-out (upper middle panel) and the Cooper-Frye freeze-out
    with the UrQMD afterburner (lower middle panel). The lowermost
    panel shows results from default UrQMD calculations without a
    hydrodynamic phase.}
  \label{fig:results-winkel-all-cells}
\end{figure}

This strong head shock, leading to a particle emission in forward
direction, can clearly be seen from the angular distribution of the
emitted nucleons in the uppermost panel of
Fig.~\ref{fig:results-winkel-all-cells}, where the hydrodynamic
scenario was applied, followed by the free streaming freeze-out.
While one clearly observes the suppression of particle emission under
small angles for $E_{\rm lab} \le 10$~AGeV, a strong head shock
develops at higher beam energies and pushes a considerable amount of
matter ahead, resulting in a particle emission at small angles
($\theta_{\rm lab} < 20^{\circ}$) and a reduced emission of particles
in $\theta_{\rm MC}$-direction. However, for all beam energies
investigated in this study, a considerable number of nucleons is
emitted at large angles ($50^{\circ} < \theta_{\rm lab} < 80^{\circ}$)
due to the sideward deflection of matter in the evolving Mach shock
wave.\par
The two middle panels in Fig.~\ref{fig:results-winkel-all-cells} show
results from hydrodynamic calculations with particle production via
the Cooper-Frye freeze-out (second panel from top) and subsequently
the UrQMD afterburner (third panel).  The maxima of the distributions
$\theta_{\rm max}$ are now located at much smaller angles within a
range $20^{\circ} < \theta_{\rm lab} < 50^{\circ}$. Here, the fully
isotropic, but Lorentz-boosted thermal particle distribution
considered in the Cooper-Frye freeze-out~\cite{Petersen:2008dd} is
superimposed on the results from the hydrodynamic calculations.\par
To understand this result one can consider a simple toy model where a
significantly strong isotropic particle distribution, boosted in beam
direction, is superimposed on the results from pure
hydrodynamics. This simple approach can reproduce the effect of the
Cooper-Frye freeze-out on the particle distribution. Note, that due to
reasons of geometry, a fully isotropic (thermal) distribution has a
$sin(\theta)$-shape in the comoving frame when plotted as
$dN/d\theta$.  If this distribution is moving with respect to the
observer (laboratory frame) the sinusodial shape is deformed and the
maximum of the distribution is shifted to smaller angles.\par
Particle rescattering and the decay of baryonic resonances within the
UrQMD afterburner, however, shift the maxima of the angular
distribution of emitted nucleons to slightly larger angles (cf.\ third
panel of Fig.~\ref{fig:results-winkel-all-cells}).\par
The lowest panel of Fig.~\ref{fig:results-winkel-all-cells} shows the
results of default UrQMD calculations. The momentum conserving
particle scattering results in much broader and smoother angular
distributions, with maxima about $30^{\circ} < \theta_{\rm lab} <
70^{\circ}$, dependent on the beam energy.  The shape of the curves
simply reflects an isotropic distribution that is Lorentz-boosted with
a certain velocity on top of the fermi distributed nucleons which did
not interact at all.\par
\begin{figure}[t]
  \centering 
  \includegraphics[width=1.\columnwidth]{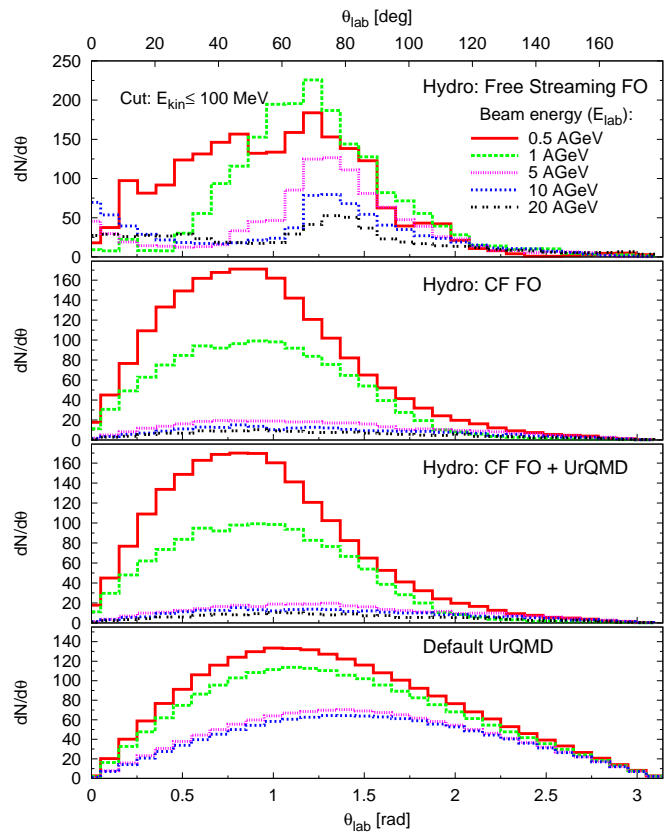}
  \caption{(Color online) Angular distributions of low energy reaction
    products ($E_{\rm kin} \le 100$~MeV) of Ne+U reactions calculated
    using the hydrodynamic approach with the free streaming freeze-out
    (upper panel), the Cooper-Frye freeze-out (upper middle panel) and
    the Cooper-Frye freeze-out with the UrQMD afterburner (lower
    middle panel). They show emission at distinct Mach cone angles in
    the investigated scenarios (cf.
    Fig.~\ref{fig:results-winkel-all-cells}). The maxima of these
    distributions $\theta_{\rm max}$ for the different scenarios are
    depicted in Fig.~\ref{fig:excitation_fct}.}
  \label{fig:results-winkel-energy-cut}
\end{figure}

Figure~\ref{fig:results-winkel-energy-cut} illustrates that a cut in
the kinetic energy per nucleon at roughly the speed of sound, which
corresponds to an $E_{\rm kin} \le 100$~MeV, changes the angular
distribution for all scenarios, in particular for the higher beam
energies ($E_{\rm lab} > 1$~AGeV). The high-energetic particles
predominantly originate from hot and dense regions of the head shock
which is explicitly excluded by the application of this
cut. Therefore, only particles emitted from the much cooler and more
dilute sideways traveling Mach shock wave are observed in this
analysis.  For the UrQMD calculations this cut removes most forward
moving participant nucleons and leaves mostly the fermi distributed
spectators in the spectrum.\par
\begin{figure}[t]
  \centering
  \includegraphics[width=1.\columnwidth]{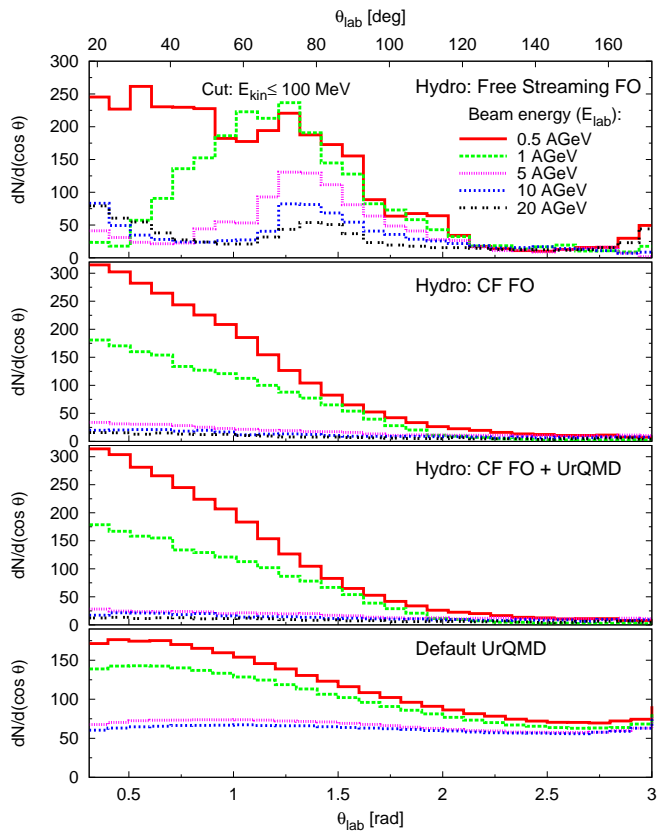}
  \caption{(Color online) Angular distribution of low energy reaction
    products (Fig.~\ref{fig:results-winkel-energy-cut}) of Ne+U
    reactions calculated using the hydrodynamic approach with the free
    streaming freeze-out (upper panel), the Cooper-Frye freeze-out
    (upper middle panel), the Cooper-Frye freeze-out with the UrQMD
    afterburner (lower middle panel), and default UrQMD (lowermost
    panel), from which the $\sin(\theta)$-shape of an isotropic
    thermal particle distribution was removed by plotting
    $dN/d\cos(\theta)$.}
  \label{fig:results-dN/dcos(theta)}
\end{figure}

Plotting $dN/d\cos\theta$ instead eliminates the sinusoidal shape as
can be seen in Fig.~\ref{fig:results-dN/dcos(theta)}. Here, the
Cooper-Frye freeze-out shows a clear peak at very small angles for
lower beam energies ($E_{\rm kin} \le 100$~MeV) due to the head shock
wave and the superimposed thermal distribution similar to the one seen
in the lowest panel.  The results from the free streaming freeze-out
(uppermost panel) and with this the Mach shock wave signal is the less
affected by this change.\par
\begin{figure}[t]
  \centering
  \includegraphics[width=.96\columnwidth]{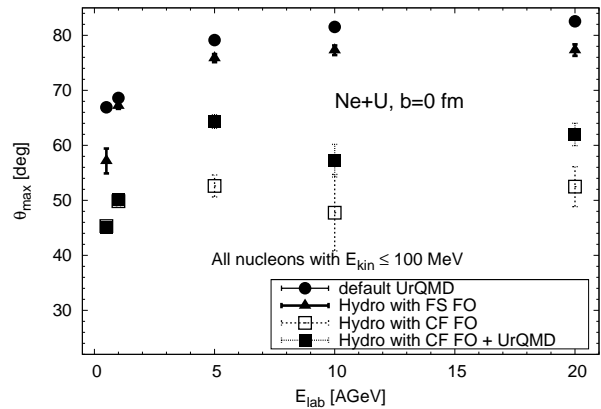}
  \caption{Beam energy dependence of the emission angle $\theta_{\rm
      max}$ for default UrQMD (full circles) and hydrodynamic
    processes, including the free streaming freeze-out (full
    triangles), the Cooper-Frye freeze-out with (full squares) and
    without (open squares) a subsequent UrQMD afterburner.}
  \label{fig:excitation_fct}
\end{figure}
Figure~\ref{fig:excitation_fct} depicts the dependence of the conical
emission angle $\theta_{\rm max}$ on the beam energy, for the
different freeze-out descriptions considered. The angles are extracted
from the underlying data of Fig.~\ref{fig:results-winkel-energy-cut}
which includes a low-energy cut so that only particles with $E_{\rm
  kin} \le 100$~MeV are considered that originate predominantly from
dilute and slow fluid cells, i.e.\ the Mach shock wave. Those did not
cross the phase transition to the chirally restored phase. Therefore,
the resulting particle emission at Mach angles is probing the purely
hadronic phase. It is obvious that the Mach wave travels with $v
\simeq c_{\rm s}$ through the hadronic phase which should result in a
rather slight change of the emission angle with varying beam
energy.\par
For the free streaming freeze-out (full triangles), the emission angle
grows slightly with increasing beam energy. For the Cooper-Frye
freeze-out disregarding the subsequent resonance decays (open squares)
however, the emission angle is notably shifted towards smaller values
due to the superimposition of the boosted thermal distribution. By
employing the UrQMD-afterburner (full squares), the computed
preferential emission angle gets shifted by about $\Delta \theta_{\rm
  lab} = +10^{\circ}$.\par
Most of the cluster-bound particles from hydrodynamics with free
streaming freeze-out are emitted at Mach angles within the range $
70^{\circ} < \theta_{\rm MC} < 80^{\circ}$ for $E_{\rm lab} >
1$~AGeV~\footnote{The emission angle in events with $E_{\rm lab} \le
  1$~AGeV is smaller than $70^{\circ}$ because the shock wave does not
  exhibit such high densities and thus moves with rather low velocity
  only [cf.\ Eq.~(\ref{eq:mach-angle})].}. This result corresponds to
a \emph{cluster flow} with the speed of sound.\par
Comparing the results obtained from the purely hydrodynamic
calculations to the results from the hadronic transport model
calculations without a hydrodynamic phase (UrQMD, full circles in
Fig.~\ref{fig:excitation_fct}) shows that even the pure transport
calculation suggests conical emission of nucleons, at angles very
similar to the expected Mach cone angles (note that, for the UrQMD
results, the same cut in kinetic energy of the nucleons is applied).
While in the hydrodynamic picture the conical emission originates from
a Mach-like wave, binary nucleon-nucleon scattering results in the
distinct particle emission pattern in the transport model which
approximates the shape of a boosted thermal distribution.  \par
It is possible to distinguish those scenarios by determining the
predominant cluster emission angle at non-vanishing impact parameters.
The left panel of Fig.~\ref{fig:angles_div_calc} shows non-central
Ne+U collisions at $E_{\rm lab}=5$~AGeV for hydrodynamic calculations
followed by the free streaming freeze-out as well as default UrQMD
calculations. Here, only nucleons emitted within the energy range of
$10$~MeV$ \le E_{\rm kin} \le 100$~MeV are taken into account in order
to ensure that nucleons originating from the spectator part of the
target nucleus are excluded. In the hydrodynamic scenario, the
extracted emission angle stays at $\theta_{\rm MC} \simeq 77^{\circ}$
up to an impact parameter of $b=4$~fm. For larger impact parameters,
$\theta_{\rm lab}$ \emph{decreases} considerably to $\theta_{\rm lab}
\simeq 60^{\circ}$ for $b=9$~fm. In this case, the projectile hits the
target at its periphery (cf.\ Fig.~\ref{fig:time_evolution_cf_energy})
and the Mach shock wave propagates through the whole of the heavy
nucleus.  UrQMD calculations however, show a different behavior. Here,
the extracted angle $\theta_{\rm max}$ \emph{grows} with increasing
impact parameter, up to $\theta_{\rm max} = 90^{\circ}$.  This
distribution pattern corresponds to a fully isotropic particle
emission.\par
Thus, the emission angle for semi-central collisions can be used to
study the reaction mechanism. While emission angles above $\theta_{\rm
  max} \simeq 80^{\circ}$ favour kinetic collision processes to cause
conical emission, emission angles well below this value indicate
collective Mach shock waves in nuclear matter.\par
The right panel of Fig.~\ref{fig:angles_div_calc} depicts the maximum
emission angle of emitted nucleons in A+U collisions with $E_{\rm kin}
\le 100$~MeV.  Here, the emission angle stays constant at $\theta_{\rm
  max} \simeq 77^{\circ}$ from $\alpha$ up to Ca projectiles. The
particle number in the low energy bin drops by about 75\%, and
collisions with Ag-projectiles show that there are virtually no
particles with kinetic energies below $100$~MeV left. The whole dense
mater system formed by the two colliding nuclei merges completely into
a highly accelerated shocked fireball~\cite{Westfall:1976fu}. Compared
to these hydrodynamic calculations, the emission angles for A+U
collisions computed with the default UrQMD model are larger by roughly
$10^{\circ}$.

\section{Summary}
We studied the emission angles $\theta_{\rm lab}$ ($\theta$ denotes
the polar angle between the beam axis and the flux of matter) of
reaction products in Ne+U collisions in a beam energy range from
$E_{\rm lab} = 0.5$~AGeV to $20$~AGeV in a hydrodynamic approach with
different freeze-out methods as well as using the UrQMD model. It is
shown that a strong head shock wave is created during the hydrodynamic
evolution due to the strong compression of nuclear matter in the
collision zone. This head shock penetrates the target nucleus with a
supersonic velocity $v_{\rm sh}$ and reaches very high energy as well
as baryon densities which are sufficient to enter the chirally
restored state of matter that is described by the EoS.\par
\label{sec:summary}
\begin{figure}[t]
  \centering
  \includegraphics[width=.96\columnwidth]{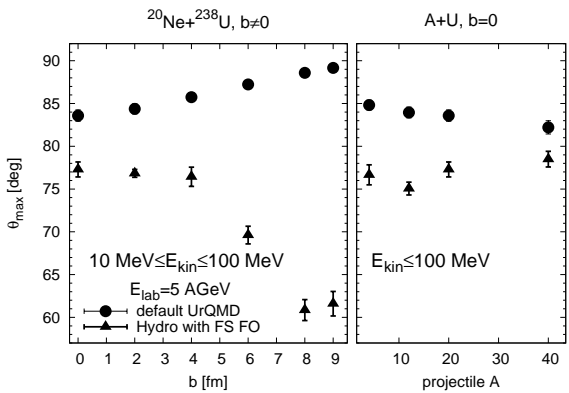}
  \caption{Maximum emission angle $\theta_{\rm max}$ of emitted
    nucleons from hydrodynamic (free-streaming freeze-out) and UrQMD
    calculations for non-central Ne+U collisions (left) and for
    central A+U collisions (right) at $E_{\rm lab}=5$~AGeV.}
  \label{fig:angles_div_calc}
\end{figure}
Behind the head shock a more dilute Mach cone like wave (\emph{Mach
  shock wave}) evolves and propagates at slower velocity ($v \simeq
c_{\rm s}$) through the hadronic state.\par
We show that the energy spectra of emitted nucleons peaks at low
kinetic energies (see Fig.~\ref{fig:results-energy-bins}). These slow
particles ($E_{\rm kin} \le 100$~MeV) are mostly emitted at Mach
angles which can be extracted from hydrodynamic calculations followed
by a free streaming freeze-out (cf.\
Figs.~\ref{fig:results-winkel-all-cells} to
\ref{fig:results-dN/dcos(theta)}). Those peaks should be visible even
for beam energies reached at the FAIR at GSI. These rather small
emission angles for all investigated beam energies lead to the
conclusion that Mach-like waves travel in the hadronic phase only.
Therefore, a crossing of the phase transition to the chirally restored
phase as it may occur in the head shock seems not to be accessible
through the emission angles of the measured particles.\par
Applying however a Cooper-Frye freeze-out (with and without the UrQMD
afterburner), these peaks are shifted to smaller angles when plotting
$dN/d\theta$ due to the superposition with the fully isotropic, but
Lorentz-boosted thermal particle distribution considered in the
Cooper-Frye freeze-out. Removing the sinusoidal shape of the boosted
thermal distribution by considering $dN/d\cos\theta$-distributions
(see again Figs.\ \ref{fig:results-winkel-all-cells} to
\ref{fig:results-dN/dcos(theta)}), these Mach cone peaks can no longer
be seen.\par
While results from standard UrQMD calculations lead to emission angles
comparable to those of a hydrodynamic calculation followed by a
Cooper-Frye freeze-out for central collisions (cf.\ Figs.\
\ref{fig:results-winkel-all-cells} to
\ref{fig:results-dN/dcos(theta)}), they show a notably different
behavior for non-vanishing impact parameters (see
Fig.~\ref{fig:angles_div_calc}, left panel).  Thus, the emission angle
can be used to distinguish the underlying process leading to a conical
emission pattern of nucleons, checking the existence of shock waves in
nuclear matter.\par
Moreover, we show that for different projectiles from $\alpha$ up to
Ca a distinct particle emission angle prevails over a wide range of
impact parameters for both hydrodynamic and UrQMD calculations (cf.\
Fig.~\ref{fig:angles_div_calc}, right panel).

\section{Acknowledgements}
\label{sec:acknowledgements}
We thank Dirk Rischke for providing the one fluid hydrodynamics code
that was used for our calculations and Giorgio Torrieri for the
fruitful discussions. This work was supported by GSI, BMBF, and by the
Hessian LOEWE initiative through the Helmholtz International Center
for FAIR (HIC for FAIR), the Helmholtz Graduate School for Heavy Ion
Research, and the Helmholtz Research School on Quark Matter
Studies. B.B.\ and H.P.\ acknowledge a Feodor Lynen fellowship of the
Alexander von Humboldt foundation. This work was partially supported
by the US-DOE Nuclear Science grant DE-FG02-93ER40764 and the U.S.\
department of Energy grant DE-FG02-05ER41367. We are grateful to the
Center for the Scientific Computing (CSC) at Frankfurt University for
providing the computational resources.

\bibliographystyle{apsrev}
\bibliography{Biblio}

\end{document}